\documentstyle[12pt]{article}
\begin{document}
\newcommand{\ot}{\frac{1}{2}}
\newcommand{\D}{& \displaystyle}  
\newcommand{\di}{\displaystyle}   
\newcommand{\abs}{\vspace*{0.5cm}\hspace*{-0.6cm}}
\newcommand{\hsp}{\hspace*{5cm}}
\newcommand{\mmath}[1]{{\mbox{\math #1}}}
\newcommand{\beh}{\stackrel{!}{=}}
\newcommand{\mbold}[1]{\mbox{\boldmath $ #1 $}}
\newcommand{\smbold}[1]{\mbox{\small$\mbold{#1}$}}
\newcommand{\hmbold}[1]{\hat{\mbold{#1}}}
\newcommand{\Cl}{\mbold{C}}
\newcommand{\CCl}{\mbold{C'}}
\newcommand{\Fl}{\mbold{F_4}}
\newcommand{\Sl}{\mbold{S}}
\newcommand{\kal}[1]{\mbox{$\cal #1 $}}
\newcommand{\OP}{\kal{O}}
\newcommand{\wG}{\widetilde G}
\newcommand{\hatphi}{\hat\Phi}
\font \math=msym10 scaled \magstep 1

\date{August 1992}
\title
{Rotational Symmetry and Regularization Dependence in the
$\Phi^4_4$-Model\thanks{Supported by Fonds zur F\"orderung der
Wissenschaftliche Forschung in \"Osterreich, project P7849.} }
\author
{\bf C.B. Lang and U. Winkler\\  \\
Institut f\"ur Theoretische Physik,\\
Universit\"at Graz, A-8010 Graz, AUSTRIA}
\maketitle
\begin{abstract}
We study the one component $\Phi^4$ model for four different lattice
actions in the Gaussian limit and for the Ising model in the broken phase.
Emphasis is put on the euclidean invariance properties of the boson
propagator. A measure of the violation of rotational symmetry serves as
a tool to compare the regularization dependence of the triviality bound.
\end{abstract}
\newpage

\section{Motivation and introduction}

There is  ample evidence that for the lattice regularized $\Phi^4$-model
the physical quartic coupling decreases logarithmically
with increasing cutoff and that the continuum limit
describes non-interacting boson fields.
Thus the scalar sector of the standard model
\begin{equation}
   S_{scalar}=\frac{1}{2}(\partial_\mu\Phi)^2+\frac{m_0^2}{2}
     |\,\Phi\,|^2+\frac{\lambda}{4!}(|\,\Phi\,|^2)^2  , \label{equ0}
\end{equation}
where the Higgs fields live in $O(4)$, has to be considered as an
effective low energy model describing the leading cutoff effects and
may be studied independently.

The cutoff then represents new, possibly more
fundamental physics which would replace the standard-model at higher
energies. However, its influence is effective on all scales and
manifests itself in the cutoff dependence of all observables.
In describing the experiments through a lattice model, the
higher in energy we go the stronger we expect such cutoff effects
to be. More and more terms would have to be included in the effective
action in order to explain experiments with sufficient accuracy.
Within the scaling region one expects to find this
dependencies to be only weak, but this knowledge is
mainly based on rough estimates and there are only very few explicit
studies \cite{BhBi88,BhBiHe90,He90,GoKaNe92a,GoKaNe92b}.

In this article we study the cutoff dependence with
emphasis on the euclidean invariance properties of the boson
propagator. As a simplified toy model we use
the one-component $\Phi^4$ model, where one does not have the additional
complication due to Goldstone modes. We expect that most of our
qualitative results carry over to $O(N)$.

One particularly interesting observable exhibiting a cutoff dependence
as discussed above is the upper bound on the effective coupling
$\lambda_R$\footnote{The subscript $R$ denotes renormalized quantities
throughout the article.}. This bound emerges at infinite bare
coupling (the Ising limit of the model) as a consequence of triviality,
and is related to the Higgs mass via the ratio
\begin{equation}
R = \sqrt{\lambda_{R} /3} = m_H/f.\label{Rbound}
\end{equation}
In first order perturbation-theory and for the $O(4)$ model
the $W$ mass may be related to the vacuum condensate
$f_\pi^{phys}\!=\!\langle\Phi_R\rangle\!\approx\!246$ GeV,
and therefore (\ref{Rbound}) provides an upper bound for the
physical Higgs mass \cite{DaNe83}. Because this quantity is essentially
nonperturbative, we need nonperturbative methods to study its cutoff
dependence.

In our further discussion we refer to different {\em lattice}
regularizations only. In lattice formulations the cut-off dependence
enters through the
discretization of the action and the geometry of the lattice mesh.
In particular we expect regularization
dependent lattice artifacts for large values of the dimensionless Higgs
mass (low values of the correlation length).
Neuberger et al.\cite{HeNeVr92,Ne92} have studied
these effects in the $O(N)$-model by systematically
adding operators of increasing order in $1/\Lambda$; they also studied
the contribution of similar higher order terms in a $O(4)$
model on an $F_4$-lattice \cite{Ne87}.

Different lattice geometries correspond to different lattice
regularizations and have been  studied mainly in
the context of the model with $O(4)$ symmetry, in connection with the
triviality bound on the Higgs mass. First results in
\cite{BhBi88} were followed by a particular careful study of the model
on the $F_4$ lattice \cite{BhBiHe90,He90}.
During completion of this article we learned of a detailed study of the
Symanzik action \cite{GoKaNe92b}.

Here we present a Monte Carlo study of the one-component
$\Phi^4_4$-model in the broken phase.
For the standard hypercubic lattice this model has
been extensively studied both, analytically \cite{LuWe87} and in various
Monte Carlo simulations \cite{phi4}.
We investigate it on altogether four different lattice geometries, introduced
in the context of the $O(4)$-model and report on a simultaneous
analysis. We compare the rotational invariance (RI) properties of the
boson propagator in the broken phase of the corresponding Ising model
with those of the related Gaussian model. We use the violation
of this invariance as a measure in order to estimate the regularization
dependence of the triviality bound.

\section{Various lattice regularizations}
\subsection{Lattice actions}

Consider the action for the one-component $\Phi^4$-model defined
on  a lattice. We study four different kinds of lattice-geometries
and corresponding actions, which we call $\Cl$, $\CCl$, $\Fl$ and
$\Sl$. All four lattices may be embedded into the regular hypercubic
one $\Lambda_L \subset {\mbox{\math Z}}^4$ of volume $L^4$, with
different couplings to nearest, next-to-nearest and on-axis distance two
neighbour spins. For convenience we present all results on this
hypercubic mesh with lattice constant $a\equiv 1$. The general action
$S$ may then be written in the form
\begin{equation}
   S=\sum_{x\in\Lambda_L}\left[-\kappa\sum_{\nu=1}^{N}\eta_\nu\hatphi_{x+
     e_\nu}\hatphi_x+\hatphi_x^2+g(\hatphi_x^2-1)^2\right]\quad .
     \label{equ1}
\end{equation}\\
By $\mbold{x}+\mbold{e}_\nu$ we denote the neighbour site in  direction
$\mbold{e}_\nu$, which may be a linear combination of the
cartesian basisvectors in 4 dimensions $\{\hmbold{e}_\mu,\mu=1,...,4\}$.

Table 1 summarizes our notation, in particular the number $N$ of
interaction neighbours, the weight factors $\eta_\nu$ and a
normalization factor $\alpha$ that has been defined in such a way that
the trivial continuum limit of the kinetic term is universal (with a
leading $p^2$ term in the inverse momentum space propagator).

We denote the `lattice field' by $\hatphi$ to distinguish it from the
field $\Phi$ in `continuum' normalization. The naive continuum action is
reconstructed in the limit $a\rightarrow 0,\,L\rightarrow\infty$
with the substitutions
\begin{eqnarray}
   m_0^2   & = & \frac{1-2g}{\alpha\kappa}-\frac{1}{\alpha}\sum_{\nu=1}
                 ^{N}\eta_\nu \nonumber \\
   \lambda & = & \frac{6g}{(\alpha\kappa)^2} \\
   \Phi_x  & = & \sqrt{2\alpha\kappa}\,\hat\Phi_x \quad  . \nonumber
\end{eqnarray}

$\Cl$ denotes the usual simple hypercubic lattice action with
interactions between nearest neighbours along the directions of
the cartesian axes. In comparison to $\Cl$ the $\CCl$-action
\cite{BhBi88}
has additional couplings to the next-to-nearest neighbours in diagonal
directions  so that there are altogether 32 interaction neighbours.
$\Fl$ is the nearest neighbour action on an
$F_4$-lattice \cite{Ne87,BhBiHe90}. It's embedding into a hypercubic
lattice may be imagined by removing all odd sites of the grid.
The symmetries for this case
forbid terms like $\sum_\mu p_\mu^4$ in the inverse momentum space
propagator and guarantee Lorentz-invariance to a higher order
in the momentum cut-off ($O(\Lambda^{-4})$) than the first two
hypercubic lattice schemes ($O(\Lambda^{-2})$). The $\Sl$ denotes the
Symanzik improved action
which  includes antiferromagnetic couplings to the the next-to-nearest
neighbours along the 4 euclidean axes-directions.
These additional terms are incorporated in order to remove the
undesired $O(\Lambda^{-2})$ contributions, like for $\Fl$.
The remaining corrections of $O(\Lambda^{-4})$ in the lattice dispersion
relation  guarantee tree-level improvement \cite{Sy83} (cf. sect. 4).

\abs
With
\begin{equation}
\kal{L}\!=\!\kal{L}_{kin}+\kal{L}_{pot}  , \quad
\kal{L}_{pot}\!=\!\hat\Phi_x^2\!+\!g\left(\hat \Phi_x^2-1\right)^2 ,
\end{equation}
the kinetic parts can be written
\begin{eqnarray}
   \kal{L}_{kin}^C     &=& -2\kappa\sum_{\mu=1}^4 \hat\Phi_{x+\hat e_\mu}
                           \hat\Phi_x \quad,\\
   \kal{L}_{kin}^{C'}  &=& -2\kappa\sum_{\mu=1}^4 \left[\hat\Phi_{x+\hat
                           e_\mu}+\sum_{\nu=1}^{\mu-1}\left(\hat\Phi_{x+
                           \hat e_\mu+\hat e_\nu}+\hat\Phi_{x+\hat e_\mu-
                           \hat e_\nu}\right)\right]\hat\Phi_x \quad,\\
   \kal{L}_{kin}^{F_4} &=& -2\kappa\sum_{\mu=1}^4 \sum_{\nu=1}^{\mu-1}\left(
                           \hat\Phi_{x+\hat e_\mu+\hat e_\nu}+\hat\Phi_{x+
                           \hat e_\mu-\hat e_\nu}\right)\hat\Phi_x\quad, \\
   \kal{L}_{kin}^{S}   &=& -2\kappa\sum_{\mu=1}^4 \left(\hat\Phi_{x+\hat
                           e_\mu}-\frac{\hat\Phi_{x+2\hat e_\mu}}{16}\right)
                           \hat\Phi_x\quad.
\end{eqnarray}
Note that  $\kal{L}_{kin}^{F_4}$ does not couple odd with even
sites.

Quantization amounts to summing over all field configurations and
taking expectation values  with regard to the
Gibbs measure $Z^{-1} \exp{(-S)}\, d\Phi$ where the partition
function $Z$ is the normalization factor such that
$\langle\mbold{1}\rangle\!=\!1$. For the
Gaussian version of the model this can be done explicitly, since the
functional integrals factorize in momentum space. For the general theory
one has to rely on expansion techniques or Monte Carlo integration.

\subsection{How to compare two regularization schemes}

Comparing results from different actions obtained
at finite cut-off is not straightforward, though. Different actions have
different scaling violation properties. Choosing couplings such that the
observed correlations lengths agree is not necessarily the best
definition. Consider e.g. an effective action for block variables (which
usually is more complicated). Comparing the results of the
original and the block action at a point in their respective coupling
constant spaces where the dimensionless correlation lengths agree, the
block action will definitely have better continuum properties.
It is ``closer to the continuum limit''.

So one has to define what one means by the intrinsic ratio between the
scale parameters in two different regularization schemes A and B.
In an asymptotically free theory that issue can be resolved by
adjusting the behaviour of the $\beta$-function in the weak coupling
region \cite{HaHa80,LuWe87,Ha89}. In our case in principle we can
rely on  renormalized perturbation theory.
In fact, the two-loop expansion gives
\begin{equation} \label{mrvsgr}
m_{phys} a_s = C_s (\beta_1 \lambda_R)^{-\beta_2/\beta_1^2}
e^{-1/\beta_1 \lambda_R} \{1 + O(\lambda_R)\}
\end{equation}
where the subscript $s$ denotes the regularization scheme (lattice
action or geometry). Asymptotically, at small enough small $\lambda_R$,
$ a_s/a_{s^\prime} = C_s /C_{s^\prime}$ can be determined by lattice
perturbation
theory (cf. the results of \cite{LuWe87} for the hypercubic lattice model).

For a comparison of different effective theories at finite cutoff and
not too small $\lambda_R$ we can no
longer rely on that approach. It would be appropriate to choose another
physical observable $\OP$, that may serve as a means to measure the
distance from the continuum limit independently. Comparing results due
to different actions $S_s$ at the same values of $\OP$ then provides an
estimate for the regularization dependence. This observable could
be for example a $\pi\pi$-scattering cross-section, the $\sigma$-decay
width or the violation of euclidean RI of some quantities
\cite{BhBiHe90,He90,GoKaNe92b}.
Different operators $\OP$ will usually have different cut-off
dependence and their choice needs further motivation: There is no
``best'' observable.
Quantities like the  ratio $R$ should be considered versus
such a ``distance'' $\OP$ to the continuum limit.
Of course we could still use the inverse correlation length
$ m_{phys} a_s$, but in order to compare different schemes we have to
rescale $a_s$ at each point such that $\OP$ agrees.

In this study for four different kinds of lattice actions we compute the
envelope of $R$ in the Ising limit ($g\!=\!\infty$). In order to relate
the respective results we use for $\OP$ the RI violations of the inverse
correlation length and the inverse 1-particle momentum space propagator.

\section{Rotational invariance (RI)}

The process of regularization by a lattice implies a violation of
rotational symmetry for non-vanishing lattice constant. Only in the
continuum limit do we hope to restore full euclidean symmetry.
Thus the amount of violation of RI provides a measure of how close one
is to the continuum.

In order to estimate the violation of RI one can in principle
refer to scattering cross sections. However, measuring these in
computer simulations is notoriously difficult. For that reason one
relied on
tree level lattice perturbative calculations in order to estimate the
leading violations for e.g. Goldstone-Goldstone scattering
in the $O(4)$ model \cite{Lu89,LuWe87,BhBiHe90,He90,GoKaNe92b} where the
Born term is proportional to the massive scalar propagator. In the one
component model this would require the study of scattering of massive
Higgs particles.

In our approach we want check the violation of RI in the Higgs
propagator \cite{La89}. That quantity is more accessible
in the simulation. For simplicity we consider only 2-d slices of our
hypercubic lattice. This has the additional advantage that the
Monte Carlo data results from operators summed over the other two
directions provide better statistics.

The problem then is to quantify the RI of some function $F$ defined
on a planar grid.
Specifying $F$ to be a real positive function in the plane, with the
symmetry property
$\!F(r,\varphi)\!\equiv\!F(r,-\varphi)\!\equiv\!F(r,\varphi\!+\
\!\frac{k}{2}\pi)$,
$k\!\in\!\mmath{Z}, \,\varphi\!\in\![-\pi,\pi)$.
Then $F$ has a series-expansion of the form
\begin{equation}
    F(r,\varphi)=\sum_{n=0}^{\infty} \alpha_n(r)\cos(4n\varphi)\quad.
\end{equation}\\
For fixed $r$ we define $\varphi_0$ to be the directional
angle for which $F$ is minimal.
For the inverse propagator our models always have
$\varphi_0=0$; for the inverse correlation length $\varphi_0=0$ for
$\Cl$, $\CCl$ and $\varphi_0=\pi/4$ for $\Fl$, $\Sl$ (in a d=2 slice).
A measure for the deviation from the RI of $F(r,\varphi)$ at the
distance $r$ can be defined by
\begin{equation}
   \kal{N}(r):=\sqrt{\frac{1}{2\pi}\int \limits_{-\pi}^{+\pi}\Bigl[\frac{
   F(r,\varphi)}{F(r,\varphi_0)}-1\Bigr]^2d\varphi}=\sqrt{\lambda_0^2(r)+
   \frac{1}{2}\sum_{n=1}^{\infty}\lambda_n^2(r)}\quad.
\end{equation}\\
where
\begin{displaymath}
   \lambda_n(r)=\frac{\alpha_n(r)}{F(r,\varphi_0)}-\delta_{n,0}\quad.
\end{displaymath}\\
The last relation follows from the orthogonality of the
cosine functions. The leading contribution to $\kal{N}$ may be estimated
from the ratio $\frac{F(r,\pi/4)}{F(r,0)}$ since in leading order
$\kal{N}\simeq \sqrt{\frac{3}{8}}\mid\frac{F(r,\pi/4)}{F(r,0)}-1\mid$.

We can estimate the violation of RI either in real space or, by
Fourier transformation, in momentum space. For infinite lattice size
the momentum variables cover finite continuous intervals.
In our discussion of the Gaussian model we utilize this property.

In a straightforward application one could study the real space
propagator for given fixed coupling $\kappa$. However, it is clear that
small and large distances (for periodic b.c.) trivially show strongest
RI violation. In
order to compare results at different values of the coupling (and the
correlation length) we have to specify a typical value of
$\mid\!\mbold{x}\!\mid$ or $\mid\!\mbold{p}\!\mid$ where we
determine $\kal{N}$ for that coupling.
We choose  $\mid\!\mbold{p}\!\mid = m(\kappa)$, a momentum value
equal to the inverse correlation length in the direction of the
cartesian axes.

For all lattice actions considered we have calculated $\kal{N}$
for
\begin{enumerate}
\item[i)] the inverse momentum-space propagator $\wG^{-1}$,
denoted by $\kal{N}_G$, and
\item[ii)] the inverse directional correlation-length $\xi^{-1}$,
denoted by $\kal{N}_\xi$ .
\end{enumerate}
In the continuum limit $m\to 0$ we have  $\kal{N}\to 0$.

We determined $\kal{N}$ for the Gaussian model both analytically
with
Taylor-expansion and by use of conventional numerical methods.
For the Monte Carlo results of our simulation of the Ising model we
relied on suitable interpolation of the numerically determined
momentum space propagator, as discussed later.

\subsubsection*{i) Inverse momentum-space propagator}

We restrict the momentum-space to two dimensions by considering
a particle which moves only in 2 of the 4 space-time dimensions with
4-momentum $\mbold{p}\!=\!(p_t,p_x,p_y\!\!=\!\!0,p_z\!\!=\!\!0)$,
with the notation $p_t\!=\!m\cos\varphi,\,p_x\!=\!m\sin\varphi$, i.e.
$\mid\!\mbold{p}\!\mid = m$.

If the propagator is known explicitly, as it is the case for the
Gaussian model, the simplest way to calculate $\kal{N}_G(m)$ is by the
series-expansion of $\wG_s^{-1}$ in the functions
$\cos(4n\varphi)$ ($s$ denotes the lattice-action). The corresponding
coefficients $\alpha_n$ and $\lambda_n$ may then be expressed through
Bessel functions and decrease rapidly.
For $n\!\ge\!1$ and $p\!\le\!1$ the ratio $\lambda_{n+1}/\lambda_n
\approx\!10^{-3n}$. Therefore it is sufficient to consider
only the first 2 coefficients \cite{La89}.
For propagator values given numerically one has to rely on an adequate
interpolation or fitting procedure.

\subsubsection*{ii) Inverse directional correlation length}

The {\it inverse directional correlation length} $\xi^{-1}_{\,s}
(m_s\,|\,\hmbold{r})$ \cite{Pa88} for the lattice-action $s$ in
direction
$\hmbold{r}=(\hat r_1,\hat r_2,\hat r_3,\hat r_4),\,\,|\,\hmbold{r}\,|\!=
\!1$, is defined implicitly by the equation
\begin{equation}
   \wG^{-1}_s(\kappa_s\,|\,\mbold{p}= i\xi^{-1}_s\hmbold{r})\beh 0\quad.
\end{equation}
In that expression $\xi^{-1}_{\,s}$ describes the asymptotic, exponential
decay of the  configuration-space propagator in direction $\hmbold{r}$.
In particular, along the main axes, e.g.
$\xi^{-1} (m|(1, 0, 0, 0))\!=\!m(\kappa)$.
Again we study the behaviour in a 2-d plane.

Even in the Gaussian model the coefficients $\alpha_n$ and
$\lambda_n$ cannot be represented exactly, but one may
express them through a power-series. The qualitative behaviour for
$m\!<\!1$ does
not differ from the one of $\wG^{-1}_s$. For numerically
given propagators one may determine the directional
correlation length from e.g. a fit along the specified direction (to the
suitably interpolated propagator). If the deviation from RI is
correctly described by the equivalent Gaussian model, one may rely to
a fit to the corresponding propagator.

\section{The Gaussian model}

Removing the quartic term in (\ref{equ1}) we are left with the
Gaussian model describing a free scalar particle. The action is a
positive-definite  quadratic form,
\begin{equation}
   S=\sum_{x,y\in\Lambda_L}\hatphi_x Q_{x,y}\hatphi_y \quad,\qquad
   Q_{x,y}=\delta_{x,y}-\kappa\sum_{\nu=1}^{N}\eta_\nu\delta_{x,y-e_\nu}\quad,
\end{equation}
and it is possible to solve explicitly for quantities like the
propagator. For the momentum-space propagator one finds
\begin{equation}
   \wG(\kappa\,|\,\mbold{p})^{-1}=2\Big[{1-\kappa\sum_{\nu=1}^{N}
       \eta_\nu\cos(\mbold{p}\!\cdot\!\mbold{e}_\nu)}\Big]\quad,
       \label{equ11}
\end{equation}
where the components of $\mbold{p}$ assume values
$\frac{2n\pi}{L}$, $n=0,\ldots,L-1$ on an $L^4$ lattice with periodic
boundary conditions. The configuration-space propagator
$\langle\Phi_y\Phi_x\rangle$
cannot be represented in similar closed analytical form. When it is
summed up over 3-space, however, it is possible to give an explicit
expression. For periodic boundary conditions
$\mbold{x}\!+\!L\hmbold{e}_\mu\!\equiv\!\mbold{x}$ and if there are
only ferromagnetic interactions (like in the cases $\Cl,\,\CCl$ and
$\Fl$), the well-known result (cf. \cite{Ro91}) is
\begin{equation}
   \sum_{\vec x}\langle\Phi_0\Phi_x\rangle=\frac{2\alpha\kappa}{L}
      \sum_{n_0=0}^{L-1}\wG\left(\kappa\,|\,p_t=\frac{2
n_0\pi}{L},\vec 0\right)
      e^{i p_t t}=\frac{e^{-mt}+e^{-m\left(L-t\right)}}
         {2(1-e^{-Lm})\sinh m} \quad.   \label{equ4}
\end{equation}
(Note that this is the propagator for fields in the continuum
notation.) The
relation between the lattice mass $m$ and the hopping parameter
$\kappa$ is defined implicitly by the pole of the momentum-space
propagator in the complex energy-plane
\begin{equation}
   \wG^{-1}\left(\kappa\,|\,im,\vec 0\right)\beh 0\quad
   \label{equ2}
\end{equation}
with the solution
\begin{equation}
\sinh{\frac{m}{2}} = \ot \sqrt{\frac{1-N \kappa}{\alpha\kappa} } \quad .
\end{equation}
This implies the critical values for $\kappa_c = \frac{1}{N}$
for these three actions (with corresponding values of $N$).

For the case $\Sl$, with  the additional antiferromagnetic interaction
term, (\ref{equ2}) assumes the form
\begin{equation}
    8 - \kappa ( 45 + 16 \cosh m -  \cosh 2m ) =0 ,     \label{equ2x}
\end{equation}
which has, for $\frac{4}{39}<\kappa<kappa_c = \frac{2}{15}$,
two real solutions $m_1$ and $m_2$ obeying the relation
$\,\,\cosh m_1\!+\!\cosh m_2\! =\!8$.
Only $m_1$ approaches zero at the critical point. The other pole
corresponds to a `ghost' with a negative residue, approaching $m_2
\to \cosh^{-1}(7)$ at the critical point. In the continuum-limit the
corresponding physical mass goes to infinity and this state decouples.

For this action (\ref{equ2}) assumes the form
\begin{eqnarray}   \label{equ5}
\sum_{\vec x}\langle\Phi_0\Phi_x\rangle=
\phantom{\sinh m_1\left[\frac{e^{-m_1t}+e^{-m_1\left(
  L-t\right)}}{(1-e^{-Lm_1})\sinh m_1}-
  \frac{e^{-m_2t}+e^{-m_2\left(L-t\right)}}{(1-e^{-Lm_2})\sinh m_2}
  \right] }&& \nonumber\\
\frac{3}{\cosh m_2\!-\!\cosh m_1}\left[\frac{e^{-m_1t}+e^{-m_1\left(
  L-t\right)}}{(1-e^{-Lm_1})\sinh m_1}-
  \frac{e^{-m_2t}+e^{-m_2\left(L-t\right)}}{(1-e^{-Lm_2})\sinh m_2}
  \right]  .&&
\end{eqnarray}
The closer to the phase transition we are, the more the physical mode
dominates the propagator.
Practically the ghost becomes noticeable only at small distances $t$.
E.g. for $m_1\!=\!1$ at $t\!=\!0$ the ghost term contributes
about 20\% to the propagator, at $t\!=\!1$ it is 4\% , but at $t=8$
its contribution has already decreased by several orders of magnitude
and may be neglected.

We now turn to the RI properties of the propagator. In the Gaussian
model we have the advantage that we can work on both, finite and
infinite volumes. We consider a two-dimensional submanifold,
either in real space or in momentum space.

Fig. 1a gives the inverse momentum space propagator
$\wG^{-1}(p,\varphi)$ for the Gaussian models and a given value of
$m=0.407$, and in Fig. 1b we plot the corresponding $\varphi$-dependence
for $p=\frac{\pi}{2}$ as an example.

Following the above discussion and requiring $\mid\mbold{p}\mid=m$, in
Fig. 2 we show lines of constant values of
$\wG^{-1}(p_x=m\sin\varphi, p_t=m\cos\varphi)$, as derived for
infinite lattice size. For small m (small lattice spacing) rotational
symmetry is restored.

All this information may be summarized by plotting $\kal{N}_G$ and
$\kal{N}_\xi$, shown in Fig. 3 for $p\!\le\!1$. In each case the RI
is violated only up to a very small amount, expecially for $\Fl$, where
$\kal{N}_G\,,\kal{N}_\xi\!<\!0.0004$ for that region of values. Even for
the standard hypercubic action $\Cl$ where the largest violations
appear, $\kal{N}$ does not exceed the value 0.012.

Generally in two or more dimensions $\xi^{-1}_{\,s}$ is only calculable
approximately, e.g. as a Taylor-series in the lattice-mass $m_s$,
\begin{equation}
\xi_s^{-1}(m_s\,|\,\hmbold{r}) =
m_s\left[1+a_s(\hmbold{r})m_s^2+O(m_s^4)\right]\quad.
\end{equation}

Both RI-functions $\kal{N}_G$
and  $\kal{N}_\xi$  may be expanded
\begin{equation}
   \kal{N}_F(m_s) = A_s\,(m_s^2)^{\gamma_s}\left[1+B_{l,F}
m_s^2+O(m_s^4)\right]\quad. \label{equ3}
\end{equation}
The leading coefficient $A_s$ is independent of $F\!\in\!\{\wG_s^{-1},
\,\xi_s^{-1}\}$, but it is not allowed to conclude that $A_s$ is a
universal
constant \cite{BhBiHe90}. The same holds for $\gamma_s$; it is a
measure for the degree of euclidean symmetry restoration in the limit
$m_s\rightarrow 0$ and has for $\Cl,\,\CCl$ the value 1 and for
$\Fl,\,\Sl$ the value 2. There are no $O(a^2)$ corrections for
$\Fl$ and the tree-level improved Symanzik action $\Sl$,
which shows the higher degree of RI for these actions,
one of the reasons for their construction.

Postulating
\begin{equation}
   \kal{N}_B(m_B)\,\beh\kal{N}_A(m_A)  \label{equ7}
\end{equation}
we could rescale the corresponding definitions of the lattice
constants and get asymptotic ratios
\begin{equation}
   \frac{a_{C'}}{a_C}  \longrightarrow \sqrt{7}\quad,\qquad
   \frac{a_{F_4}}{a_S} \longrightarrow \root 4 \of{6} \quad,
\end{equation}
for the Gaussian models.

\section{Results for the Ising limit}

The upper bound on the renormalized coupling and on the Higgs boson mass
is realized for bare coupling $g \to \infty$, the Ising limit. Thus it
has become customary to study $\Phi^4$-models in that simple limit
restricting the range of values of the field variable to $\pm 1$.

In our Monte Carlo simulation we worked on lattices of size  $16^4$ and
periodic boundary conditions. Each of the 4 actions was considered at
$4-6$ values of the coupling
parameter $\kappa\!>\!\kappa_{cr}$ in the broken phase, close to the
critical point $\kappa_{cr}$ such that $m\!\in\!(0.2,0.9)$.
For the updating we used the cluster algorithm \cite{SwWa87}
(actions $\Cl,\,\CCl$ and $\Fl$) or the Metropolis method (action
$\Sl$, where we had problems with the cluster updating due to presence
of additional antiferromagnetic interaction terms). In the
determination of the propagators we utilized reduced variance operators
\cite{MoMuWo88}. For each data point we considered $40\,000$ (cluster
algorithm) or $100\,000$ (Metropolis) update-measurement pairs, in
addition to a suitable number of equilibrating updates.

In this way we obtained the boson propagator summed over $y,\,z$
directions
\begin{equation}
G(\kappa|t,x)\!:=\!\langle\Phi_{0,0,0,0}\sum_{y,z}\Phi_{t,x,y,z}\rangle
\end{equation}
and its Fourier transform. We also determined the  susceptibility
\begin{equation}
\chi  = \sum_{t,x} G(\kappa|t,x) = \frac{Z_R}{m_R^2}\quad.
\label{susceptibility}
\end{equation}
In order to extract the required quantities $m_s\!=\!a_s m_{phys}$ and
$R$, we used (\ref{equ4}) and (\ref{equ5}) for the
Gaussian propagator
\begin{equation}
   G_s(t):=\sum_{x}G_s(t,x)\approx a(e^{-mt}+e^{-m(L-t)})+b\quad,
\label{propfit}
\end{equation}
where $b$ gives the square of the vacuum expectation value,
$\langle\Phi\rangle^2$. For our analysis we therefore use this value
for $\mid\!\langle\Phi\rangle\!\mid$.  The direct estimate of
$\langle\Phi\rangle$ is not possible due to tunneling between the two
ground states.

For the tree level improved action  $\Sl$ we also add another term to
(\ref{propfit}),
\begin{equation}
a' (e^{-m'(m)t}+e^{-m'(m)(L-t)})\quad ,
\end{equation}
where $m'(m)$ is assumed to be given by the relation $\cosh m\!+\!\cosh
m'\!=\!8$ which is valid for the corresponding Gaussian model.

The maximal relative statistical errors for the fit parameters amount to
$\Delta m\approx\pm 2\!-\!3\%,\,\Delta a \approx\pm 1\!-\!2 \%,\,\Delta
b\approx\pm 0.1\%$ for results due to cluster updating and measuring.
For the conventional updating and measurement (case $\Sl$) the errors are
typically larger by a factor 4.

In Fig. 4  we plot $\kappa_I$ vs. $\kappa_G$ such that the masses
determined for the Ising- and the Gaussian model at
corresponding values of the reduced coupling
$\tau = \mid\!\kappa -\kappa_c\!\mid /\kappa_c$ agree. On an
infinite lattice one expects $\tau_G \propto \tau_I \mid\!\ln
\tau_I\!\mid^{-\frac{1}{3}}$ and the extrapolation to the critical
values $\kappa_{G,c}$ thus should provide an
estimate of $\kappa_{I,c}$. On finite size however, the mass does
not vanish at the phase transition. Still, within the accuracy of our
results the points obey the scaling law quite accurately and
we confirm that the leading critical exponents of $m(\tau)$ agree in
the two cases; the logarithmic factor cannot be resolved.

Comparing the RI properties of our Ising model
results for $G(t,x)$ with those of the Gaussian version for same mass we
find surprising agreement (cf. \cite{La89}). In Fig.s 1a and 1b we
compare the inverse Fourier transformed propagator $\wG^{-1}(p,\varphi)$
with the results due to the Gaussian model, to point out the
good agreement. Indeed, within the error bounds we could not
find any systematic deviations from the Gaussian behaviour. This
conforms with the expectation that the boson propagator in the broken
phase of the Ising model is in leading order that of a free particle
like in the Gaussian model -- up to logarithmic corrections.
But for the rotational symmetry properties of the propagator and the
considered quantities $\kal{N}$ those seem to be small enough to
permit neglecting them.
This justifies to use the Gaussian values as our measure for violation
of RI at given mass, at least for the mass values considered here.
This is what we do in the subsequent discussion.

In Table 2 we summarize our results for mass, $f = \langle\Phi_R\rangle
= \langle\Phi\rangle /\sqrt{Z}$, wave function renormalization constant
determined from (\ref{susceptibility}), the ratio $R$ of (\ref{Rbound})
and the value of $\kal{N}_G$, one of our measures of violation of
rotational symmetry of the boson propagator. The large errors for the
results $\Sl$ are due to the different updating method used there. In
Fig.s 5 and 6 we plot  $m_R$ and $f$ vs. the coupling and compare with
the fit to the expected asymptotic behaviour  \cite{BrGuZi76,LuWe87} for
$\tau\to 0$
\begin{eqnarray} \label{mbehaviour}
m_R &\sim& C_4 t^{\ot} \mid\!\ln t\!\mid^{-\frac{1}{6}} ,\nonumber \\
f &\sim & C_ft^{\ot} \mid\!\ln t\!\mid^{\frac{1}{3}} .
\end{eqnarray}
In Tab. 3 we give the results of our joint fit to the data for $m_R$ and $f$.
The ratios $C_4/C_f$ are compatible with the asymptotic
value $4\pi\sqrt{2}/3$ \cite{LuWe87}.

Fig. 7 gives $R$ vs. $1/m_R$, where one could in principle read
off the triviality bound. Comparing the results
for actions $\Cl$ and $\Sl$ we find a difference similar to that
observed in the recent study of $O(4)$ models in \cite{GoKaNe92b}.
However, as discussed, it may be more appropriate to use another
quantity than
$1/m_R$, in order to compare the action (and regularization)
dependence of that bound. We return to that point later.

{}From these data one may obtain the constant $C'$ (cf. \cite{LuWe87})
relating $m_R$ to the renormalized coupling,
\begin{equation} \label{asymptotic}
C' = \lim_{\lambda_R\to 0} \left[ m_R\, (\beta_1
\lambda_R)^{\beta_2/\beta_1^2} e^{1/\beta_1 \lambda_R}\right] .
\end{equation}
Depending on the closeness to the continuum limit, this fit to the
asymptotic behaviour may be not very reliable. Also, $C'$ is
determined just from our data for lattice size $16^4$ and should be
extrapolated to infinite size in a finite size scaling study.
Within these limitations we obtain the numbers  given in Tab. 3.
For action $\Cl$ L\"uscher and Weisz\cite{LuWe87} determined the
asymptotic value of $C=5.3(1.2)$; this is definitely larger
than ours. However, fitting their results in the region of mass values
considered here, we get a value of $\simeq 3.0(3)$. We are obviously not
asymptotic enough for $\Cl$.

As discussed it may not be wise to compare the
values of $R$ for different actions with each other at some
lattice mass $m_R$. Instead we follow \cite{BhBiHe90,GoKaNe92b} and
plot in Fig. 8 the renormalized coupling vs. $\kal{N}_G$, our
measure for the amount of violation of RI.

First we notice, that the difference between the results for $\Cl$ and
$\CCl$ and  between  those for $\Fl$ and $\Sl$ appears to become smaller,
as compared to the presentation in Fig. 7; it is typically less than
10\%. However, the difference between $\Cl$, $\CCl$ on one hand and $\Fl$,
$\Sl$ on the other appears to be sizeable, about 30-40\%.

In the results for the $O(4)$ model for $\Fl$ \cite{BhBiHe90} and $\Sl$
\cite{GoKaNe92b} the authors estimated the regularization dependence by
comparing the results at given value of another measure of RI, derived
from a perturbation expansion of the Goldstone-Goldstone scattering
cross-section. There the difference between $\Fl$,  $\Sl$ compared to
$\Cl$ was smaller than ours, of the order of 10-20 \% for 1-10\%
violation of RI. Although this variation could be due to the
study of different models (one component vs. four component
fields) this may also indicate a strong dependence on the choice of
measure for RI. Maybe a systematic study of higher order terms, like
the one initiated recently \cite{HeNeVr92} helps to clarify this
issue.

Let us summarize our results for the Ising model:
\begin{itemize}
\item
The similarities in the propagator between the Gaussian version and the
Ising version within the same  type of action (regularization) are
bigger than between Gauss $\leftrightarrow$ Gauss or between
Ising $\leftrightarrow$ Ising for different lattice actions.
Therefore the RI behaviour for the discussed range of couplings seems to
be dominated mainly by leading kinematical effects, which are already
given in the corresponding Gaussian model.
\item
The difference between the two measures of RI, $\kal{N}_G$ and
$\kal{N}_\xi$ are less than 1\%. This result is due to the close
relationship between the correlation length and the
momentum space propagator.
\item
The regularization dependence of $R$ comparing
$\Cl$ with $\CCl$ and $\Fl$ with $\Sl$ is less than
10\% in the considered domain of couplings.
\item
The regularization dependence of $R$ comparing
$\Cl$ or $\CCl$ with $\Fl$ or $\Sl$ is about 30-40 \% in the considered
domain of couplings.
\end{itemize}

The last mentioned result allows two interpretations:
The first is, that the regularization dependence of the envelope indeed
may be about 40 \%. A similar conclusion was put forward in
\cite{HeKlNe92,HeNeVr92}.
On the other hand the studies \cite{BhBiHe90,GoKaNe92b} for the $O(4)$
model do not support this strong variation.
So the second interpretation may be that there is substantial
sensitivity on the definition of a measure of RI.

Leaving aside the issue of the regularization dependence
of the triviality bound, we found that different lattice actions lead
to propagators with remarkable rotational symmetry, quite
contrary to the naive expectation.
The leading violations are well described already by the corresponding
Gaussian models.
\subsubsection*{Acknowledgements}
We wish to thank U. Heller, J. Jers\'ak and T. Trappenberg for discussions.

\newpage

\newpage
\section*{Tables}

\noindent Table 1:
The parameters for the four actions considered, defined by
embedding into the hypercubic lattice.
\vspace{.3cm}
\begin{center}
\begin{tabular}{|c|cccl|}
\hline
$\qquad l$    & $\quad N$  & $\eta_\nu$  &  $\alpha  $  &  $\quad\{\mbold{e}
                                                           _\nu,\,\nu=1,...,N
                                                           \}$\\
\hline
$\qquad \Cl$  & $\quad 8$  & 1          &   $ 1$ & $\quad\pm\hmbold{e}_\mu$\\
$\qquad \CCl$ & $\quad 32$ & 1          &   $ 7$ & $\quad\pm\hmbold{e}_\mu,\,
                                                   \pm(\hmbold{e}_\mu\pm
                                                   \hmbold{e}_\lambda),\,
                                                   \lambda<\mu$\\
$\qquad \Fl$  & $\quad 24$ & 1          &   $ 6$ & $\quad\pm (\hmbold{e}_\mu
                                                   \pm\hmbold{e}_\lambda),\,
                                                   \lambda<\mu$\\
$\qquad \Sl$  & $\quad 16$ & $\eta_{1\!-\!8}\!\!=\!\!1,\,\eta_{9\!-\!16}\!\!=
                             \!\!-1/16$ & $ 3/4$ & $\quad\pm \hmbold{e}_\mu,
                                                   \,\pm 2\hmbold{e}_\mu$\\
\hline
\end{tabular}
\end{center}

\vspace{0.5cm}
\noindent  Table 2:
The parameters for our results in the broken phase of the
Ising limit.
\vspace{.3cm}
\begin{center}
   \begin{tabular}{|c|l|lllll|}
   \hline
   \multicolumn{1}{|c|}{Action} &
   \multicolumn{1}{c|}{$\kappa$} &
   \multicolumn{1}{c}{$m_R$} &
   \multicolumn{1}{c}{$f\equiv\langle\Phi_R\rangle$} &
   \multicolumn{1}{c}{$Z_R$} &
   \multicolumn{1}{c}{$R$} &
   \multicolumn{1}{c|}{$\kal{N}_G$} \\
   \hline
     $\Cl$  & 0.1507 & 0.244(4) & 0.0832(1) & 0.929(2) & 2.93(4) & 7.6(2)E-4 \\
            & 0.1513 & 0.328(3) & 0.1043(6) & 0.924(4) & 2.14(4) & 1.37(2)E-3\\
            & 0.1520 & 0.407(3) & 0.1227(5) & 0.918(6) & 2.31(3) & 2.09(3)E-3\\
            & 0.1525 & 0.449(1) & 0.1339(2) & 0.909(3) & 2.35(1) & 2.55(1)E-3\\
            & 0.1550 & 0.637(2) & 0.1769(1) & 0.886(2) & 2.60(1) & 5.07(3)E-3\\
            & 0.1560 & 0.698(5) & 0.1914(3) & 0.873(3) & 2.65(3) & 6.06(8)E-3\\
     \hline
     $\CCl$ & 0.0336 & 0.231(1) & 0.0896(1) & 0.974(2) & 2.58(1) & 9.80(9)E-5\\
            & 0.0338 & 0.319(2) & 0.1198(2) & 0.965(2) & 2.66(2) & 1.88(9)E-4\\
            & 0.0340 & 0.393(6) & 0.1415(3) & 0.964(5) & 2.78(5) & 2.88(9)E-4\\
            & 0.0346 & 0.560(10)& 0.1903(3) & 0.955(3) & 2.95(4) & 6.0(2)E-4 \\
            & 0.0354 & 0.738(4) & 0.2372(4) & 0.944(3) & 3.11(2) & 1.07(1)E-3\\
     \hline
     $\Fl$  & 0.0460 & 0.255(3) & 0.0875(1) & 0.959(2) & 2.91(3) & 1.79(8)E-6\\
            & 0.0470 & 0.474(1) & 0.1460(1) & 0.936(1) & 3.27(1) & 2.12(2)E-5\\
            & 0.0482 & 0.670(2) & 0.1873(2) & 0.930(2) & 3.58(1) & 8.34(9)E-5\\
            & 0.0502 & 0.917(4) & 0.2368(6) & 0.913(4) & 3.87(2) & 2.85(5)E-4\\
     \hline
     $\Sl$  & 0.1653 & 0.23(2)  & 0.078(2)  & 0.88(4)  & 3.0(4)  & 0.7(2)E-5 \\
            & 0.1660 & 0.33(1)  & 0.100(2)  & 0.89(2)  & 3.3(2)  & 3.0(4)E-5 \\
            & 0.1675 & 0.47(2)  & 0.134(2)  & 0.85(3)  & 3.5(2)  & 1.2(2)E-4 \\
            & 0.1683 & 0.55(3)  & 0.147(4)  & 0.85(4)  & 3.7(3)  & 2.3(5)E-4 \\
            & 0.1694 & 0.61(2)  & 0.165(2)  & 0.82(2)  & 3.7(1)  & 3.4(4)E-4 \\
     \hline
   \end{tabular}
\end{center}

\vspace{0.5cm}
\noindent  Table 3:
Results for $\kappa_c^{(16)}$, $C_4$ and $C_f$ from
a joint fit to $m_R(\tau)$ and $f(\tau)$ according to
(\ref{mbehaviour}); we also give $C'$
according to a fit to (\ref{asymptotic}).
\vspace{.3cm}
\begin{center}
\begin{tabular}{|c|cccc|}
\hline
Action   & $\kappa_c^{(16)}$&  $C_4$  &  $C_f$  &  $C'$   \\
\hline
 $\Cl$       & 0.14984(2)&  4.24(1)  &  0.636(1)  &  2.9(1)  \\
 $\CCl$      & 0.03340(1)&  3.76(1)  &  0.674(1)  &  6.4(2) \\
 $\Fl$       & 0.04559(1)&  3.33(1)  &  0.550(1)  &  3.2(1)  \\
 $\Sl$       & 0.16447(8)&  4.38(8)  &  0.624(5)  &  2.4(1)  \\
\hline
\end{tabular}
\end{center}

\newpage
\section*{Figures}

{\noindent \bf Fig. 1:~}
(a) The inverse momentum space propagator
$\wG^{-1}(p,\varphi)$ for the Gaussian model (action type \Cl)
for $m=0.407$; the points give the numeric results
from a simulation of the corresponding Ising model at coupling
$0.1520$ such that $m$ agrees (cf. Table 2).
(b) The values of $\wG^{-1}(p=\pi/2,\varphi)$ compared to the
corresponding values of the Ising model simulation (resulting from an
interpolation of points along the cartesian directions on the $16^4$
lattice grid).

\vspace{12pt}
{\noindent \bf Fig. 2:~}
Lines of constant values of $\xi^{-1}(m,\varphi)$, as derived in the
Gaussian model for infinite lattice size, for the four actions
considered. For small m (small lattice spacing) rotational symmetry is
restored.

\vspace{12pt}
{\noindent \bf Fig. 3:~}
Our measures for the violation of RI
$\kal{N}_G$ (full lines) and  $\kal{N}_\xi$ (dashed lines
for $m\!\le\!1$.

\vspace{12pt}
{\noindent \bf Fig. 4:~}
We plot $\kappa_I$ vs. $\kappa_G$ such that the masses determined at
corresponding values of $\tau$ agree (action $\Cl$; triangles:
$L=8$, squares: $L=16$, circles: results for
infinite size due to \cite{LuWe87}).

\vspace{12pt}
{\noindent \bf Fig. 5:~}
$f$ vs. $\kappa$ for the four actions considered; the curves give the
fits to the expected critical behaviour (\ref{mbehaviour}).
All data is for $L=16$; in the figure for $\Cl$ the triangles denote the
results for infinite lattice size due to \cite{LuWe87}.

\vspace{12pt}
{\noindent \bf Fig. 6:~}
$m_R$ vs. $\kappa$ for the four actions considered; the curves give the
fits to the expected critical behaviour (\ref{mbehaviour}).
All data is for $L=16$; in the figure for $\Cl$ the triangles denote the
results for infinite lattice size due to \cite{LuWe87}.

\vspace{12pt}
{\noindent \bf Fig. 7:~}
$R$ vs. $1/m_R$ for the four actions considered (bars: $\Cl$, squares:
$\CCl$, triangles: $\Fl$, asterisks: $\Sl$); the
curves give the fits to the asymptotic behaviour (\ref{asymptotic}).
All data is for $L=16$.

\vspace{12pt}
{\noindent \bf Fig. 8:~}
The ratio $R$ vs. $\kal{N}_G$; the curves denote fits to the
asymptotic behaviour $R(m_R)$, where $m_R$ is related to the
corresponding value of $\kal{N}_G$ as discussed. The left-hand part
gives the results on an extended scale. Symbols are like in Fig. 7.

\newpage

\end{document}